# Theoretical prediction of two-dimensional CrOF sheet as a ferromagnetic semiconductor or a half-metal

Tiantian Xiao, Guo Wang* and Yi Liao*

*Department of Chemistry, Capital Normal University, Beijing 100048, China.*

*Email:* wangguo@mail.cnu.edu.cn, yliao@cnu.edu.cn

**ABSTRACT** Two-dimensional chromium oxide fluoride CrOF sheet was studied based on density functional theory. The investigation indicates that the CrOF sheet is an intrinsic ferromagnetic semiconductor. The calculated low cleavage energy implies that the ferromagnetic semiconductor can be exfoliated from its bulk form. The corresponding Curie temperature is 150 K. In particular, the Curie temperature increases up to 410 K under hole doping and the CrOF sheet becomes a half-metal. The versatile electronic and magnetic properties indicate that the two-dimensional CrOF sheet can be a promising candidate for next-generation spintronic devices.

## 1. Introduction

Spintronics is one of the most promising next generation information technologies in the 21st century, which uses the spins of electrons as information carriers. Being different from conventional electronics, spintronics possesses potential advantages of high speed, high circuit integration density, and low power consumption [1]. Spintronic materials, such as magnetic metal and magnetic semiconductors, can serve as electrodes and the central region of spintronic devices [2]. In order to minimize the size of spintronic devices, low-dimensional ferromagnetic (FM) materials are eagerly needed. However, most of the available low-dimensional materials in the pristine form are intrinsically nonmagnetic or antiferromagnetic (AFM). Although defect, composition engineering or the proximity effect can introduce magnetic responses locally or extrinsically, intrinsic magnetism is necessary [3]. In the last year, two-dimensional $Cr_2Ge_2Te_6$ and $CrI_3$ sheets with long-range FM ordering were fabricated [3,4]. Nevertheless, low Curie temperatures limit their applications.

Hence, from the theoretical aspect, it is important to design and find intrinsic FM

materials from low-dimensional structures [5-7]. Since the discovery of graphene and its fascinating properties [8], a large number of low-dimensional materials have emerged such as boron nitride, silicon carbide, organometallic molecular wires, and single-layer transition metal compounds and so on [9]. Layered transition metal compounds constitute a large family of materials as important candidates for next generation electronic devices due to their versatile properties. For instance, $CrBr_3$ is the first FM semiconductor found in 1960, with Curie temperature of only 37 K [10]. When exfoliated from their bulk forms, the Curie temperatures of the two-dimensional chromium trichlorides $CrX_3$ (X = F, Cl, Br and I) [11,12] sheets are much increased. The Curie temperature of the two-dimensional $CrI_3$ was predicted to be 95 [11] or 107 K [12]. In a very recent study, FM CrOCl and CrOBr sheets were investigated. Being different from the $CrX_3$ that can be exfoliated from their FM bulk forms, the CrOCl and CrOBr have AFM bulk forms. Excitingly, the Curie temperature of CrOCl exceeds the record of the most-studied dilute magnetic GaMnAs materials [13]. For applications without expensive low temperature equipments, the Curie temperature should be at the room temperature or even higher due to the heat generated during operations. In this work, a two-dimensional CrOF sheet was studied theoretically. The investigation indicates that the Curie temperature of the CrOF sheet exceeds those of CrOCl and CrOBr calculated with the same method. Moreover, the calculated Curie temperature 410 K is much higher than the room temperature when hole doping is introduced.

## 2. Computational details

The geometrical optimization and electronic properties calculations were carried out with the Perdew-Burke-Ernzerhof (PBE) [14] density functional in the framework of generalized gradient approximation (GGA). The projector-augmented wave potential with an energy cutoff 400 eV included in the Vienna ab initio simulation package [15] was used. The GGA+U method was used to deal with the strongly correlated 3d electrons of Cr atoms. The value of the difference between the onsite coulomb and exchange parameter was set to 3.5 eV [16]. The density of the Monkhorst-Pack k-point mesh is about 0.05 Å$^{-1}$. For the two-dimensional sheet, a supercell with a vacuum space of 15 Å along the z direction was used in order to avoid possible interaction between images. To accurately describe the properties of solids, the screened hybrid HSE06 functional [17] that includes a portion of accurate exchange was also used. The phonon dispersion was calculated under density functional perturbation theory with the aid of the Phonopy code [18].

## 3. Results and discussions

As shown in Fig. 1(a), there are two Cr layers in the two-dimensional CrOF sheet. They are passivated by outer O and F atoms. Each primitive cell contains two Cr, two O, and two F atoms. In order to confirm the ground state, both FM and AFM configurations were considered in the calculations. In the FM configurations, the spins on different Cr atoms are parallel to each other. After geometrical optimization, the Cr-Cr distance between neighbor cells are 2.96 and 3.87 Å along the x and y directions, respectively. The nearest distance between two Cr atoms in different layers is 3.04 Å. For other cases, the Cr-Cr distances are larger than 4.88 Å. For this reason, three types of couplings were considered in this work. As shown in Fig. 1(b)-1(d), interlayer AFM couplings between Cr atoms exist in the AFM1 configuration, while there are AFM couplings between Cr atoms along the x or y direction in the AFM2 or AFM3 configuration. In the calculations, only the primitive cell is needed for the FM and AFM1 configurations. A 2×1 or 1×2 supercell was constructed for calculating the AFM2 or AFM3 configuration. The relative energies per primitive cell of the geometrically optimized structures with the FM, AFM1, AFM2, and AFM3 configurations are 0, 44, 35, and 63 meV, respectively. The energy difference should come mainly from the magnetic couplings because the geometries change little for different configurations. The calculations indicate that the FM configuration is the ground state of the CrOF sheet. The total magnetic moment per primitive cell is 6 $\mu_B$. The local magnetic moment on a Cr, O or F atom is 3.06, -0.07, and 0.01$\mu_B$. The magnetism is mainly located on the Cr atoms.

The band structures of the FM configurations are shown in Fig. 2(a). Those for the majority and minority spins are different due to the non-zero total magnetic moment. Direct band gap exists for the majority spin while indirect band gap exists for the minority spin. The band gaps for the two types of spins are 2.33 and 5.26 eV, respectively. The two-dimensional CrOF sheet is an FM semiconductor. The band structures are further confirmed by the calculations with the HSE06 density functional, which can usually better describe the band structures of solids [19]. As shown in Fig. 2(b), the main change is the larger band gaps. The direct and indirect band gaps characteristics for the majority and minority spins as well as the band shape have little change. The total magnetic moment is the same with that calculated with the PBE density functional.

To confirm the stability of the two-dimensional free-standing CrOF sheet, phonon dispersion was calculated. A 5×5 supercell was used. As shown in Fig. 3(a), no nontrivial imaginary frequency was found. There are only three imaginary frequencies near the Γ point. Detailed analysis indicates that these small imaginary frequencies (no more than 7 cm$^{-1}$) are related to translational modes. This suggests that the CrOF sheet is dynamically stable. Meanwhile, in order to examine the thermal stability of the CrOF sheet, ab initio molecular dynamics simulation was performed. In this calculation, a 4×3 unit cell was used so that the lattice parameters are larger than 10 Å. The simulation temperature was set to 300K. The step size is 1 fs. The variation of the energy with respect to the simulation time was plotted in Fig. 4(a). The energy oscillates with the time for many times, implying that equilibrium was achieved. As shown in the inset, no essential structure deformation was found after the 5000 simulation steps. These suggest that the CrOF sheet is thermally stable at room temperature.

Low cleavage energy is essential for two-dimensional materials in order to obtain successful exfoliation. For this reason, the three-dimensional bulk CrOF was calculated. Like bulk CrOCl and CrOBr, bulk CrOF has a layered structure with weak interaction between halogen atoms in adjacent layers. Because the van der Waals forces play a very important role in interlayer binding, the optPBE-vdW [20,21] density functional was used to calculate the three-dimensional CrOF. The two-dimensional one was also calculated with the same density functional for comparison. Besides the above FM and three AFM configurations, another AFM configuration AFM4 was considered, in which the Cr-Cr atoms couples ferromagnetically inside the two-dimensional layer while antiferromagnetically along the z direction. The relative energies of the bulk structures with the FM, AFM1, AFM2, and AFM4 configurations are 0, 243, 29, and -2 meV, respectively. The AFM3 configuration was not obtained though various initial structures were used. The energy -2 meV of the AFM4 configuration is quite close to that of the FM configuration. The minus value indicates that the AFM4 configuration is the ground state. Thus the two-dimensional FM CrOF sheet is exfoliated from the AFM bulk structure. With respect to the energy of this configuration, the cleavage energy of the CrOF sheet is calculated to be 0.68 $Jm^{-2}$, which is comparable to the value for graphene (0.36 $Jm^{-2}$) [22]. The cleavage energy is slightly larger than those for the two-dimensional CrOCl and CrOBr [13] but is smaller than the value 1.14 $Jm^{-2}$ for the two-dimensional $GeP_3$ [23]. The low cleavage energy should be related to the stacking pattern. In the optimized geometry, the stacking is through the F layers that belong to different CrOF units. There is no covalent bond between layers. These indicate that exfoliation from bulk structure can be an applicable method for obtaining the CrOF sheet.

The two-dimensional CrOF sheet possesses low cleavage energy. Meanwhile, it is dynamically and thermally stable. These suggest that the CrOF sheet could be realized experimentally. Next, the Curie temperature of the FM CrOF sheet is investigated. Monte Carlo simulations with an 80×80 supercell were performed based on the Heisenberg model

$$H = -\sum_{i,j} J_1 S_i \cdot S_j - \sum_{k,l} J_2 S_k \cdot S_l - \sum_{m,n} J_3 S_m \cdot S_n$$

where $J$ is the exchange parameter and $S$ is the spin of a Cr atom. In this supercell, the number of the Cr atom is larger than $10^4$. The simulation step is $10^8$. Before the simulation, exchange parameters should be calculated from the Hamiltonian. Because the magnetism is mainly on the Cr atom, the $S$ for a Cr atom is set to 3/2. According to the AFM configurations shown in Fig. 1(b)-1(d), the energy functionals are expressed as follows:

$$E_{FM} = -\frac{9}{2}J_1 - \frac{9}{2}J_2 - 9J_3 + E_0$$

$$E_{AFM1} = -\frac{9}{2}J_1 - \frac{9}{2}J_2 + 9J_3 + E_0$$

$$E_{AFM2} = \frac{9}{2}J_1 - \frac{9}{2}J_2 + E_0$$

$$E_{AFM3} = -\frac{9}{2}J_1 + \frac{9}{2}J_2 + E_0$$

Then the exchange parameters can be obtained as:

$$J_3 = (E_{AFM1} - E_{FM})/18$$

$$J_1 = (E_{AFM2} - E_{FM})/9 - J_3$$

$$J_2 = (E_{AFM3} - E_{FM})/9 - J_3$$

For comparison, the exchange parameters of the two-dimensional CrOCl and CrOBr sheets are also calculated with the same method. It should be noted that the two-dimensional CrOI sheet was also calculated. The relative energies of the structures with the FM, AFM1, AFM2, and AFM3 configurations are 0, -70, -89, and -38 meV. Thus the two-dimensional CrOI sheet has an AFM2 ground state and was not discussed here.

The three exchange parameters of the two-dimensional CrOF sheet listed in Table 1 are all positive, indicating FM couplings between Cr atoms. Among the three types of couplings, $J_2$ is the largest. From Fig. 1(a), it can be seen that $J_2$ corresponds to the coupling between adjacent Cr atoms along the y direction. The smallest $J_1$ implies that the FM coupling between adjacent Cr atoms along the x direction is the weakest. Detailed geometrical analysis indicates that two neighbor Cr atoms along the x direction are bridged by an O and a F atoms with Cr-O and Cr-F bond lengths of 2.02 and 1.96 Å, respectively. Along the y direction, the two Cr atoms are intermediated by an O atom with a bond length of 1.98 Å. The O and F atoms should affect the coupling between the Cr atoms.

In order to get insight into the bonding characteristics, the projected density of states is shown in Fig. 5. The d orbitals of the Cr atoms, the p orbitals of the O and F atoms contribute mainly to the total density of states near the Fermi level. The peaks of the projected and total density of states appear at the same energy windows. This indicates that bonds are formed between these orbitals. From the spin density shown in the inset of Fig. 5, it can be seen that the magnetism is mainly located on the Cr atoms. There is also small magnetism on the O and F atoms. Being different from the F atoms, the spins on the O atoms are opposite to those on the Cr atoms. In the two-dimensional CrOF sheet, the Cr atoms are hexa-coordinated. Crystal field theory can be applied to analysis the magnetic couplings. Although the six ligands are not equal, the five d orbitals of a Cr atom can be roughly divided into three $t_{2g}$ orbitals and two $e_g$ orbitals. There are three d electrons in a $Cr^{3+}$ ion. Thus the $t_{2g}$ orbitals with lower energy are half-occupied. Direct FM coupling between two Cr atoms is forbidden because of Pauli exclusion principle. Since the O atoms have opposite spins, indirect coupling between two Cr atoms mediated by an O atom should be feasible. Although the nearest distance between two Cr atoms along the y direction (3.87 Å) is larger than that along the x direction (2.96 Å), the $J_2$ is much larger than $J_1$. The reason is that the coupling along the y direction is mediated by an O atom while the coupling along the x direction is by an O and a F atom. The O atoms with opposite spins are favorable to the FM coupling while the F atoms with the same spins are not.

For comparison, the exchange parameters of the two-dimensional CrOCl and CrOBr sheets were also calculated. The exchange parameters are -0.95, 6.09, and 2.96 meV for the CrOCl sheet while 0.61, 4.37, and 0.94 meV for the CrOBr sheets. For the two sheets, $J_2$ is also the largest while $J_1$ is the smallest. The absolute values of the two $J_1$ for the two sheets are less than 1 meV, indicating that the coupling along the x direction is also the weakest. The minus value of $J_1$ for CrOCl indicates weak AFM coupling. After Monte Carlo simulations, the Curie temperatures can be obtained. As shown in Fig. 4(b), the Curie temperature of the two-dimensional CrOF sheet is 150 K, which is much higher than the 80 and 90 K for the two-dimensional CrOCl and CrOBr sheets. The calculated values for the two-dimensional CrOCl and CrOBr sheets are smaller than the 160 and 129 K reported in ref. [13], in which only one exchange parameter was considered. Although the Curie temperature of the two-dimensional CrOF sheet is higher than that of the two-dimensional CrI$_3$ [11,12], it does not reach room temperature.

Because direct FM couplings between two Cr$^{3+}$ ions are quasi-forbidden described above, carrier doping should be a feasible way to elevate the coupling strength. For this reason, the structure of the two-dimensional CrOF sheet with carrier doping were also optimized. The doping amounts are 0.25, 0.5, 0.75, and 1 per primitive cell. The calculations indicate that when doped with electrons, the AFM configurations become the ground states. On the contrary, the FM configurations are always the ground states with hole doping. As shown in Fig. 4(b), the Curie temperature increases with the hole doping amount. A doping with 0.25 hole per primitive cell elevates the Curie temperature to 310 K, which is much higher than that of the undoped structure and is slightly higher than the room temperature. This value is close to the 323, 314, and 293 K for the two-dimensional CrCl$_3$, CrBr$_3$, and CrI$_3$, respectively [12]. Usually, devices are heated at work. Higher Curie temperatures are needed. When the doping amount increases to 0.5, 0.75, and 1 hole per primitive cell, the Curie temperature increases to 410, 560, and 670 K, respectively. These are more than 100 ℃ higher than the room temperature and should be much useful for real applications. It is noted that a doping density of $2\times10^{15}$ cm$^{-2}$ has been achieved in experiment using electrolyte as a gate dielectric [24]. Considering the lattice parameters, the doping densities are 2.2, 4.4, 6.6, and $8.7\times10^{14}$ cm$^{-2}$ in the above four cases for the two-dimensional CrOF sheet. Thus the doping densities are within a reasonable amount.

In order to confirm the doped structures, phonon dispersions were calculated. The structures are still stable when the hole doping amount is no more than 0.50 per primitive cell. The phonon dispersion shown in Fig. 3(b) is quite similar to that of the undoped structure and no nontrivial imaginary frequency was found. Further doping introduces significant imaginary frequencies. Thus the Curie temperature 410 K can be achieved based on the stable structure. The hole doping can be applied with a gate voltage. After doping, the two-dimensional CrOF sheet becomes a half-metal. With the increasing hole doping amount, more valence bands for the majority spin are unoccupied. This can be seen in Fig. 2(c) and 2(d).

## 4. Conclusions

Structure, electronic and magnetic properties of the two-dimensional CrOF sheet were investigated based on the density functional theory. The calculated cleavage energy is 0.68 Jm$^{-2}$, indicating the exfoliation from bulk structure is possible. The CrOF sheet is an intrinsic FM semiconductor. Phonon dispersion calculations and ab initio molecular dynamics simulations imply that the CrOF sheet is stable. Direct and indirect band gaps exist for the majority and minority spin. Detailed analysis indicates that the d orbitals of the Cr atoms as well as the p orbitals of the O and F atoms contribute mainly to the total density of states near the Fermi level. The intermediated O atoms with opposite spins are favorable to the FM couplings between Cr atoms. The Curie temperature calculated by the Monte Carlo simulation is 150 K, higher than those of the two-dimensional CrOCl and CrOBr calculated with the same method. Moreover, the Curie temperature can be increased to 310, 410, 560, and 670 K when doped with hole amount of 0.25, 0.5, 0.75, and 1 per primitive celll. After doping, the two-dimensional CrOF sheet becomes a half-metal. Phonon dispersion indicates that the structure is stable when the doping amount is no more than 0.5. Thus the Curie temperature 410 K can be achieved based on the stable structure. This value is more than 100 ℃ higher than the room temperature. The two-dimensional CrOF sheet is a promising candidate for future low-dimensional spintronics.

Table 1 Exchange parameters of the two-dimensional CrOF sheet.

| hole amount | $J_1$ | $J_2$ | $J_3$ |
|---|---|---|---|
| 0 | 1.43 | 4.52 | 2.46 |
| 0.25 | 1.26 | 6.72 | 6.04 |
| 0.5 | 2.80 | 16.26 | 6.15 |
| 0.75 | 5.32 | 25.55 | 6.20 |
| 1 | 7.03 | 31.58 | 4.33 |

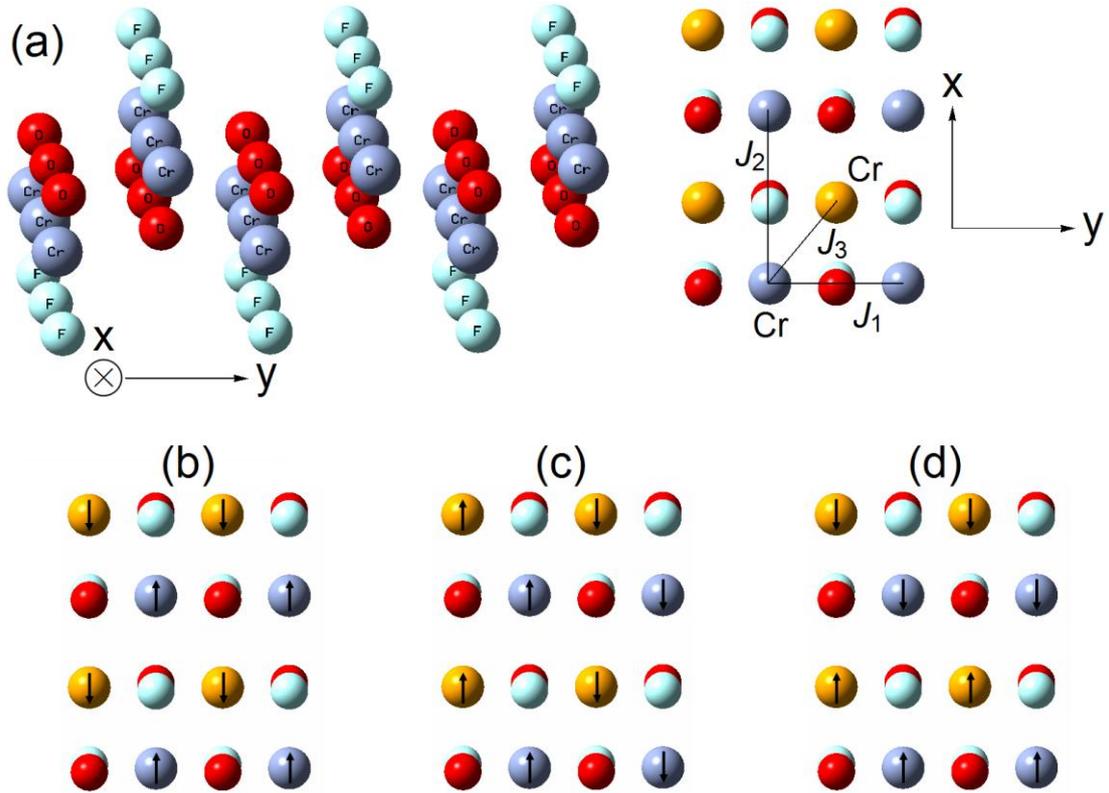

**Fig. 1.** (a) Top and side view of two-dimensional CrOF sheet. (b) AFM1, (c) AFM2, and (d) AFM3 configurations. In side views, Cr atoms in different layers are marked with different colors.

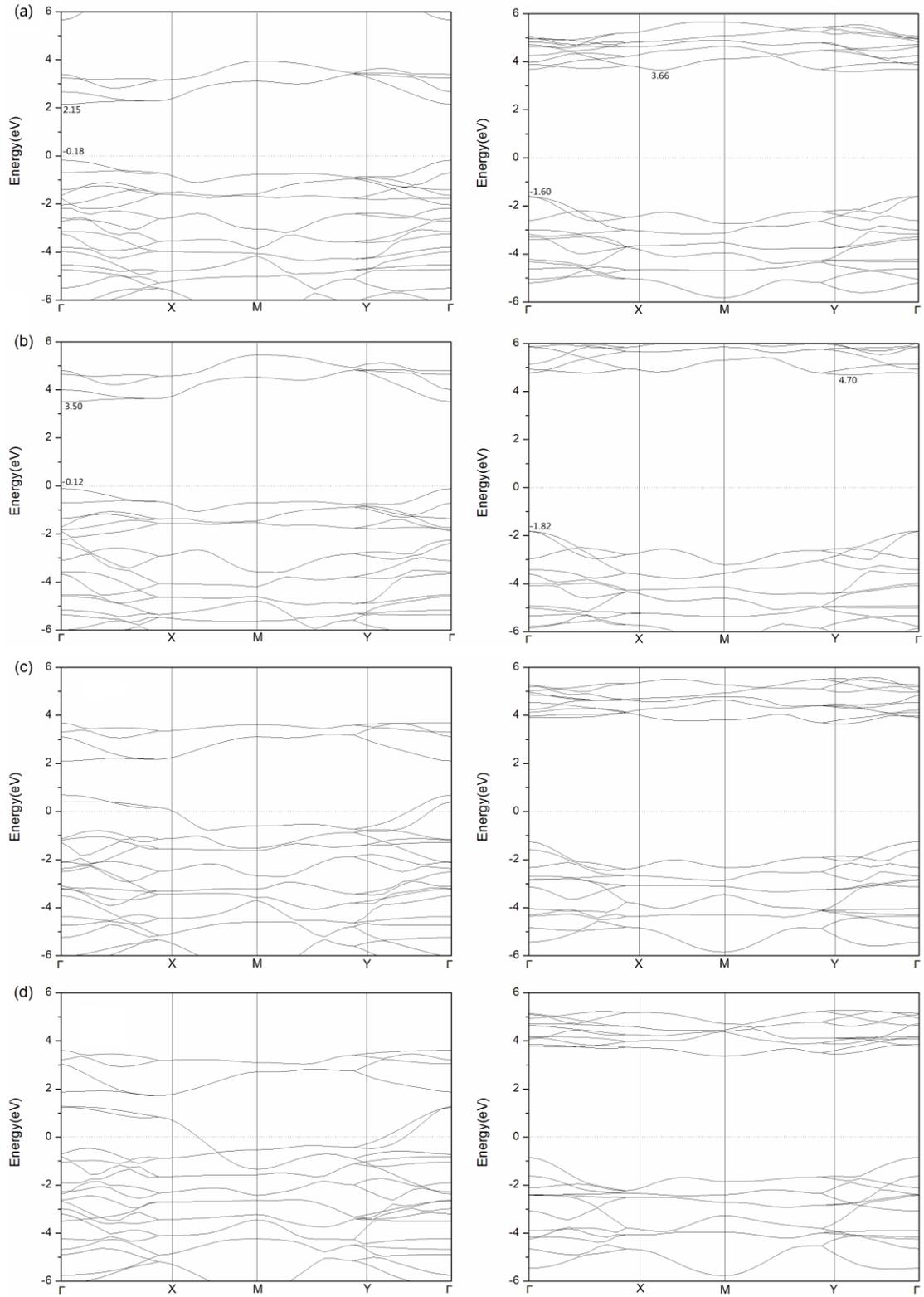

**Fig. 2.** Band structures of two-dimensional FM CrOF sheet calculated with (a) the PBE and (b) HSE06 density functional. Band structures of two-dimensional FM CrOF sheet doped with (c) 0.5 and (d) 1 hole per primitive cell. The left is for the majority spin and the right is for the minority spin. The Fermi levels are set to zero.

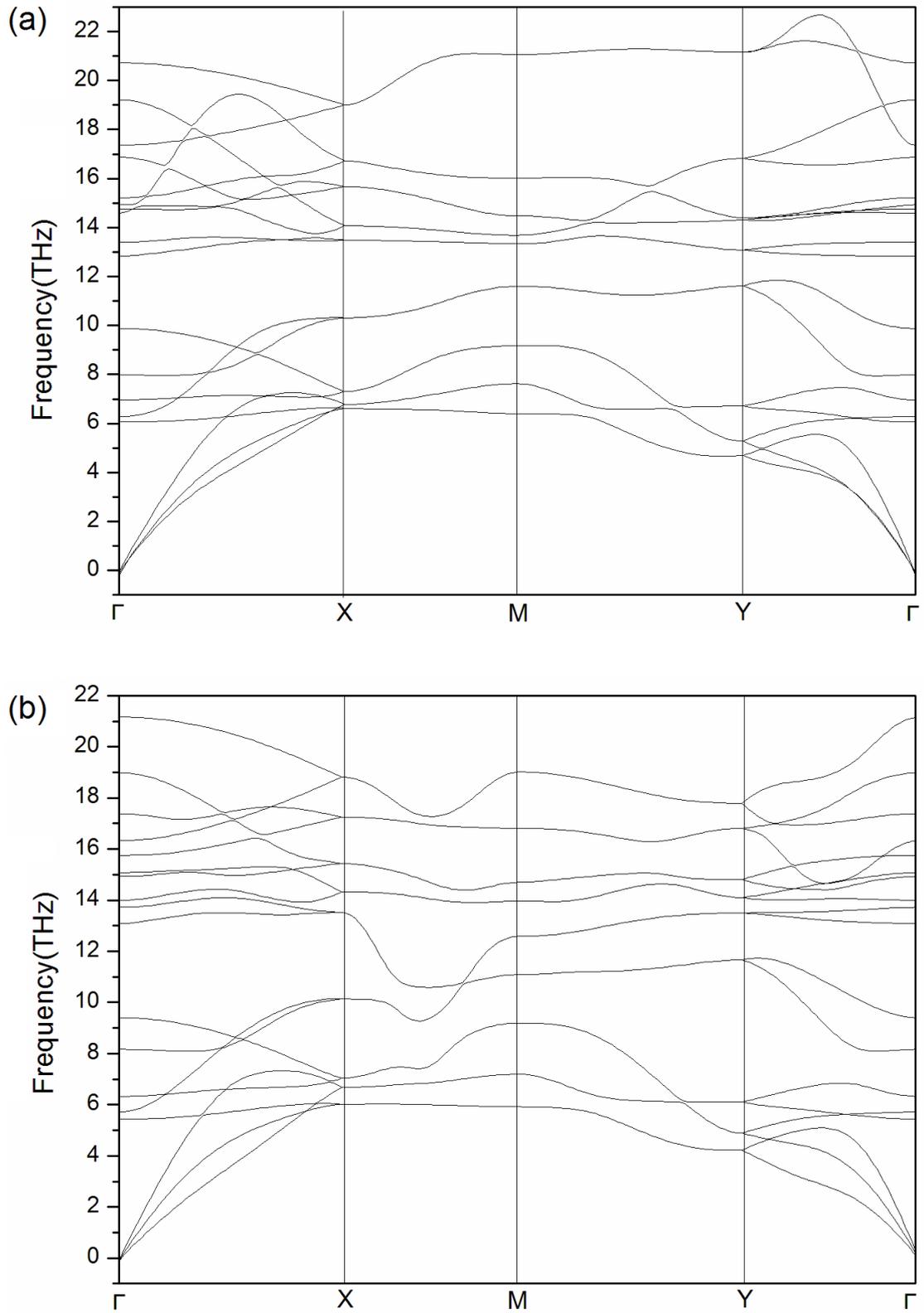

**Fig. 3.** Phonon dispersions of two-dimensional FM CrOF sheet (a) without and (b) with a hole doping amount of 0.5 per primitive cell.

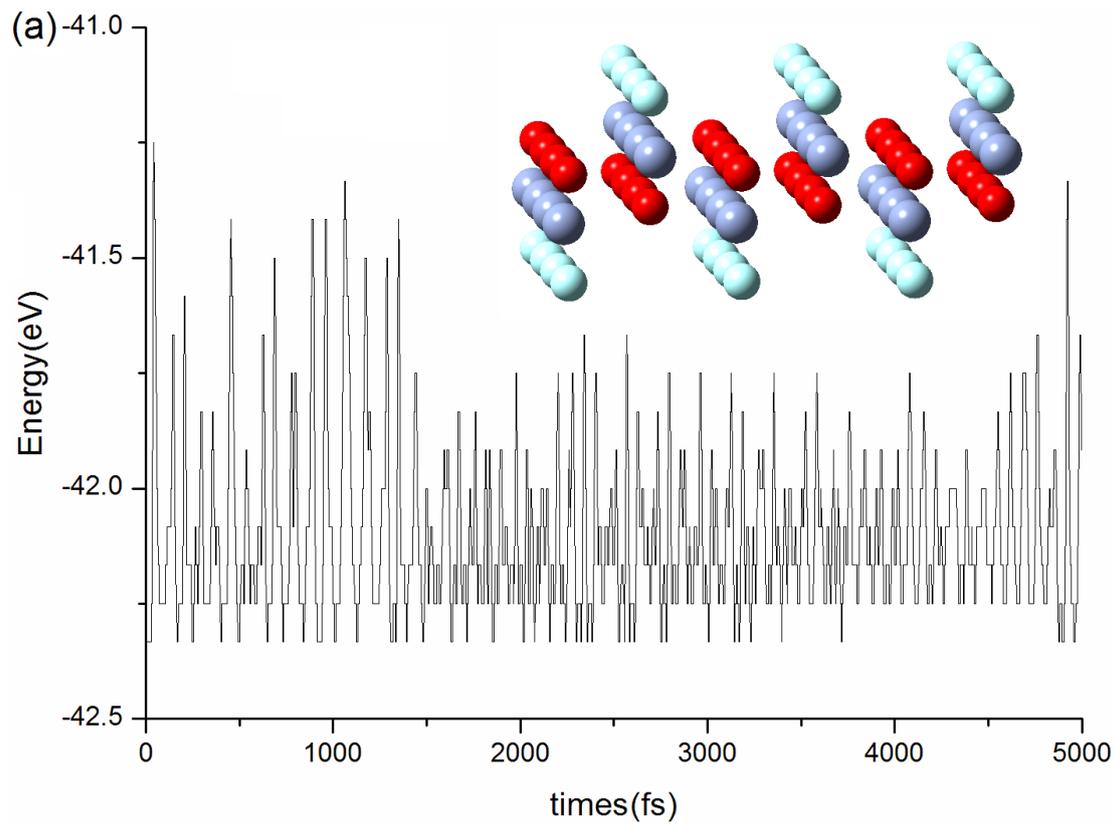

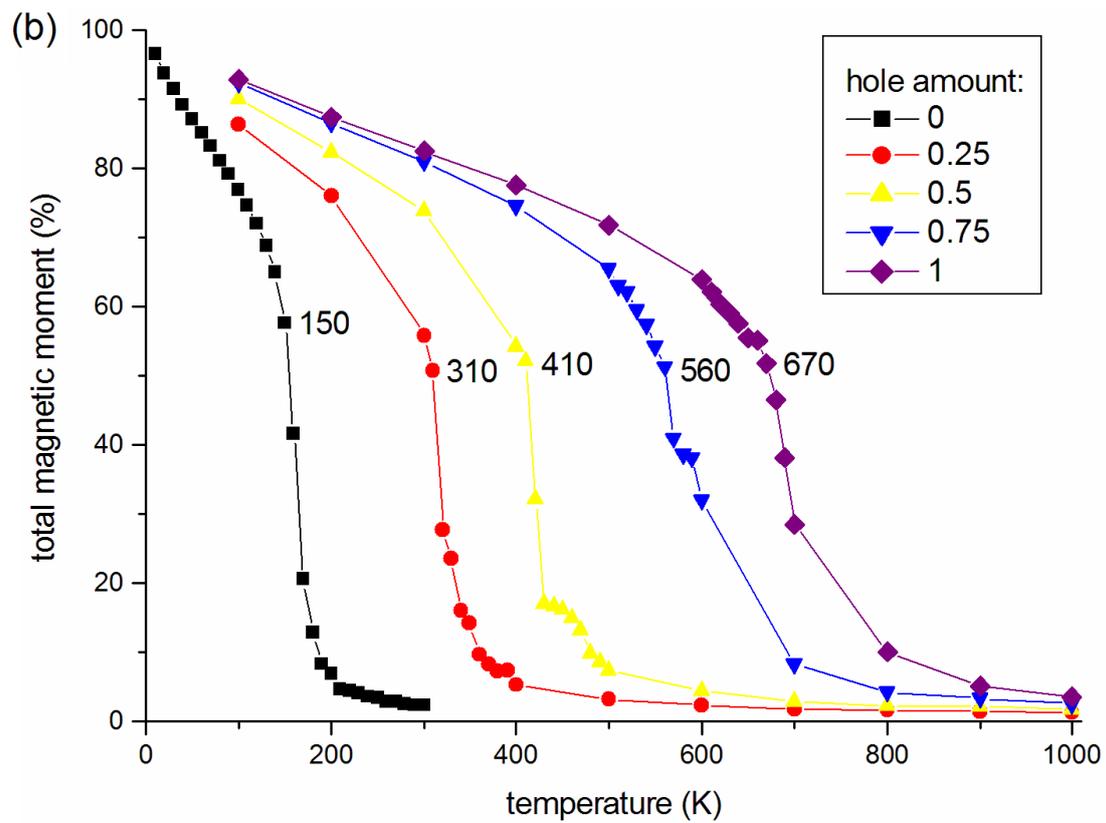

**Fig. 4.** (a) Molecular dynamics and (b) Monte Carlo simulations of the two-dimensional FM CrOF. sheet.

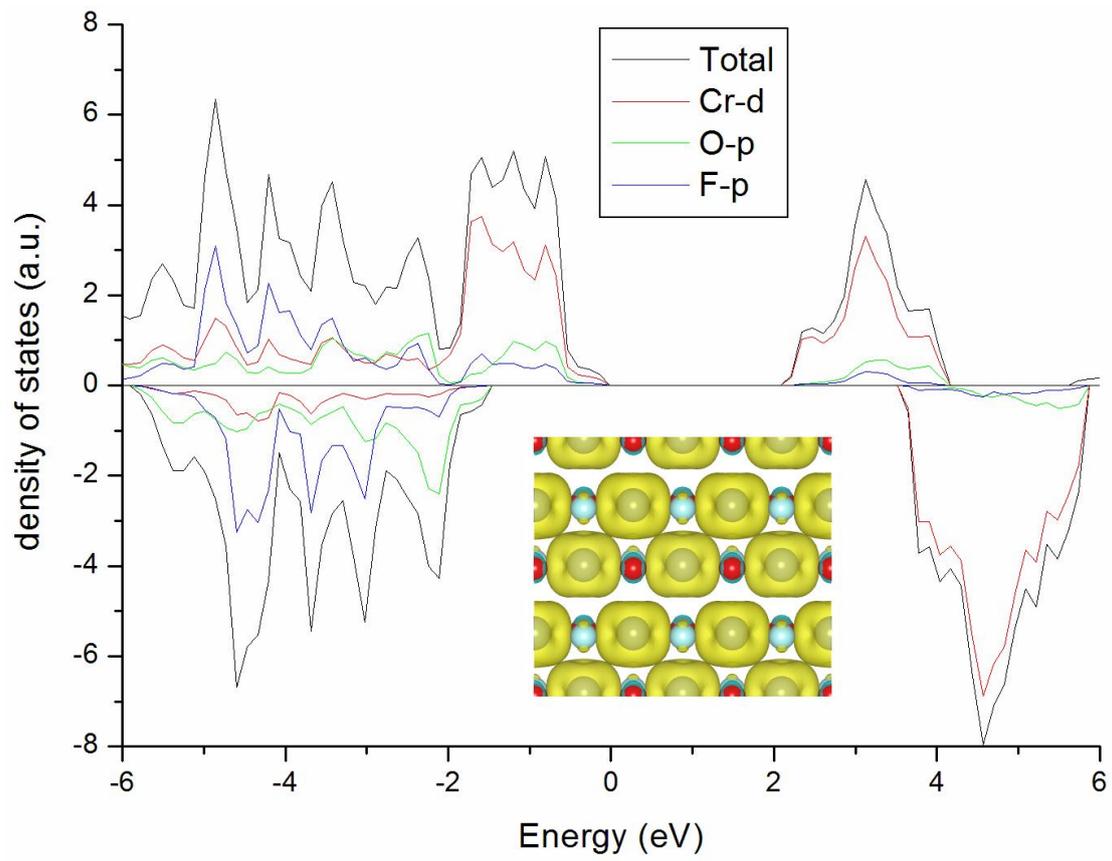

**Fig. 5.** Density of states of the two-dimensional FM CrOF sheet, the inset is the spin density.